\documentstyle[11pt]{article}

\begin{document}
\newcommand{\ECM}{\em Departament d'Estructura i Constituents de la
Mat\`eria
                  \\ Facultat de F\'\i sica, Universitat de Barcelona \\
                     Diagonal 647, E-08028 Barcelona, Spain \\
                                      and           \\
                                    I. F. A. E.        \\
                     Universitat Aut\`onoma de Barcelona \\
                     E-08193 Bellaterra, Spain }

\def\thefootnote{\fnsymbol{footnote}}
\pagestyle{empty}
{\hfill \parbox{6cm}{\begin{center} hep-ph/9710xxx\\
                                    October 1997
                     \end{center}}}
\vspace{1.5cm}

\begin{center}
\large{$\eta - \eta'$ mixing from $U(3)_L \otimes U(3)_R$ Chiral 
Perturbation Theory}

\vskip .6truein
\centerline {P. Herrera-Sikl\'ody\footnote{e-mail: herrera@ecm.ub.es},
J.I. Latorre\footnote{e-mail: latorre@ecm.ub.es},
P. Pascual\footnote{e-mail: pascual@ecm.ub.es}
and J. Taron\footnote{e-mail: taron@ecm.ub.es}}
\end{center}
\vspace{.3cm}
\begin{center}
\ECM
\end{center}
\vspace{1.5cm}

\centerline{\bf Abstract}
\medskip
We  obtain explicit expressions
for the $\eta$ and $\eta'$ masses and decay constants using
$U_L(3) \otimes U_R(3)$ chiral perturbation theory at next-to-leading
order in a combined expansion in $p^2$ and $1/N_c$. A numerical fit of 
the parameters appearing to this order of the expansion is also discussed.

\newpage
\pagestyle{plain}

The masses of the I=0 pseudoscalar $\eta$, $\eta'$ particles
as well as the value of their mixing angle have long been the subject of
discussion and theoretical interest, also in recent years. 
Phenomenologically, the situation of the $\eta - \eta'$ mixing remains not 
completely settled,
the reason beeing the high sensitivity of the mixing angle $\theta$ on
the deviation ($\Delta$) of the $\eta_8$\footnote{$\eta_0$, $\eta_8$
are the eigenstates of $SU(3)_{R+L}$. Although in the text
we are occasionally loose when referring to $\eta'$, $\eta$ instead of 
$\eta_0$, $\eta_8$, the distinction is carefully made in the calculation.}
mass from the Gell Mann-Okubo ($\Delta=0$) relation \cite{D}
\begin{equation}
m_{\eta_8}^2= \frac {1}{3} ( 4 m_K^2 - m_{\pi}^2) (1 + \Delta). 
\label{E1}
\end{equation}
Indeed, if $\Delta=0$ the singlet-octet mixing yields $\theta \sim -10^o$,
whereas data on $J/\Psi \to \eta (\eta') \gamma$ and $\eta (\eta', \pi^0)
\to \gamma \gamma$ favour a higher mixing $\theta \sim -20^o$
\cite{GK}, a factor of two bigger, that can be accomodated
from (\ref{E1}) with just a small $\Delta \sim 0.16$, 
(see also ref. \cite{BE}, which advocates for a smaller mixing).

The value of the mixing angle is linked to the value of the
$\eta'$ mass, which is heavier than the octet of light pseudoscalars,
$M_{\eta'} = 958\  {\rm MeV} \sim 2 M_K$, but still lighter than the next
I=0 candidate $\eta(1279)$, which is so distant from the $\eta$ mass that
it might not significantly mix with it. Therefore, we only 
consider $\eta - \eta'$ mixing.

Within the $SU(3)_L \otimes SU(3)_R$ Chiral Lagrangian \cite{GL} analysis,
the shift downwards of the $\eta$ mass due to this mixing is an
$O(p^4)$ effect of $SU(3)_{R+L}$ breaking,
proportional to $(m_s - \frac{m_u + m_d}{2})^2$ and
multiplied by the constant $L^{[SU]}_7$, which contains information
of the $\eta'$ state that has been integrated out.

\vspace*{0.5cm}

In this article we re-analyse these issues on the 
basis of large-$N_c$ Chiral Perturbation Theory ($\chi$PT). The $\eta'$
is regarded as the ninth goldstone boson of a nonet, together
with the octet of light pseudoscalars, and the corrections are treated
in a double expansion, both
in powers of quark masses $m_{quark}$ and $1/N_c$. Such a formulation
for this problem is plausible because $M_{\eta'}^2 - M_{octet}^2$ is an
effect induced by the $U_A(1)$ anomaly, which vanishes in the large-$N_c$ 
limit as $1/N_c$. Furthermore,
the fact that phenomenollogically it is found that 
$M_{\eta'} - M_{octet} \sim M_{octet}$, together with the usual bookkeeping 
of quark masses as $m_{quark} \sim O(M_{octet}^2)$, suggests that
the relative magnitude of the double expansion may be
regarded as 
\begin{equation}
m_q \sim 1/N_c \sim {M^2_{octet}} \sim p^2 \sim O(\delta);
\label{E2}
\end{equation}
this is the approach that we adopt and henceforth we shall refer to it
as the {\it combined expansion}. It has already been
used in the literature \cite{L}, 
in particular Leutwyler has forcefully pursued 
the analysis of  the masses of $\eta$ and $\eta'$ in order to discard the 
possibility that $m_u =0$. In this note we reproduce Leutwyler's results and
proceed further to obtain a fit for the masses and decay constants
of the goldstone boson nonet \cite{pdb96}.

Let us start from the $U_L(3)\otimes U_R(3)$ chiral lagrangian 
which has been recently proposed to $O(p^4)$ \cite{Nos}, from which we 
share the
notation and conventions, and which we refer to for references therein.
The nonet of fields are gathered in a unitary 
$3 \times 3$ matrix $U \in U(3)$, whose determinant 
$\det U= \exp (i \sqrt{6} \eta^0 /f)$ differs from unity by the presence of the
$\eta^0$ field. The chiral lagrangian ${\cal L}_{(p^0)} + {\cal L}_{(p^2)} 
+ ...$ reads
\begin{eqnarray}
{\cal L}_{(p^0)} &=& -W_0(X)\ , \nonumber\\
{\cal L}_{(p^2)} &=& W_1(X) \langle D_\mu U^\dagger D_\mu U\rangle
+W_2(X) \langle  U^\dagger \chi +\chi^\dagger U\rangle
+ i W_3(X) \langle  U^\dagger \chi -\chi^\dagger U\rangle \nonumber \\
&+&  W_5(X) \langle U^\dagger (D_\mu U)\rangle  \langle 
i a_\mu \rangle.  \nonumber\\
\label {lagrangian}
\end{eqnarray}
We have only kept the axial source because we take the divergence
of the axial current as the interpolating fields for the nonet.
The axial source $a_\mu$ 
appears in the last term 
as well as through the covariant derivatives of the fields:
\begin{eqnarray}
D_\mu U &=& {\partial}_\mu U -\frac{i}{2} \left( a_\mu U +
U  a_\mu \right). \nonumber\
\end{eqnarray}
The quark masses appear in
\begin{equation}
\chi= 2\ B\ {\rm diag}(m_u,m_d,m_s)\ .
\label{E4}
\end{equation}
As a consequence of the $U_A(1)$ anomaly, the lagrangian has
coefficient functions that depend on $X= \log (\det U )=i \sqrt{6} \eta^0 /f$,
which are not fixed by symmetry arguments. From the large-$N_c$
perspective however, only the lowest powers of $X$ survive, for each 
new power of $X$ is suppressed by a factor of $1/N_c$. Of interest to
us, the only terms that prevail are: 
\begin{eqnarray}
W_0(X) &=&  {\rm Constant} \ + \frac {f^2}{4} v_{02} X^2 + ... 
\qquad ,\qquad
\;\;\;\;\; W_1(X)= \frac {f^2}{4}+ ...  \ ,\nonumber\\
W_2(X)&=& \frac {f^2}{4}+ ... \qquad ,\qquad
\;\;\;\;\;\;\;\;\;\;\;\;\;\;\;\;\;\;\;\;\;\;\;\;\;\;\;\;\; 
W_3(X) = -i \frac {f^2}{4} v_{31} X
+ ... \ , \nonumber\\
W_5(X)&=& \frac {f^2}{4}v_{50}+... \ ,\nonumber\\
\label{E6}
\end{eqnarray}
with $v_{02}$, $v_{31}$, $v_{50} \sim O(1/N_c)$.
We shall expand the chiral lagrangian and keep terms up to $O(\delta^2)$,
according to the combined power counting (\ref{E2}). 
Recall that $f^2 \sim O(N_c)$, $B \sim O(1)$.
Within this approximation the 
chiral logarithms are suppressed (they appear at $O(\delta^3)$), although
some terms of the same order $O(p^4)$ have to be included at tree level.
More precisely, the contributions
\begin{equation}
{\cal L}_{(p^4)}= L_5 \langle D_\mu U^\dagger \; D^\mu U \;
\left( U^\dagger \chi + \chi^\dagger U \right) \rangle +
L_8 \langle \chi^\dagger U \chi ^\dagger U + U^\dagger \chi U^\dagger \chi
\rangle 
\label{E5}
\end{equation}
should be included, since 
$L_5$ and $L_8$\footnote{In 
the $U(3)_L \otimes  U(3)_R$
chiral lagrangian, instead of constants the $L_i$'s are functions of
the $\eta_0$ field. When we write $L_5$ or $L_8$ 
throughout the article we
mean the value of those functions with the argument set to zero,
$L_5 (0)$, $L_8 (0)$. They are not to be confused
with the constants in \cite{GL}, for which we reserve a superscript of $[SU]$
$L^{[SU]}_i$.}  are $O(N_c)$.
Notice that a term $L_7 \langle
U^\dagger \chi - \chi^\dagger U \rangle^2$ appears at $O(\delta^4)$
- $L_7$ is $O(1)$ in $1/N_c$ (see \cite{dRP}). The rest of terms all remain 
subleading.

The masses and the decay constants can be obtained from
the two-point function of axial currents, which is most easily done
by taking functional derivatives with respect to $a_\mu$. We shall thus
keep only terms quadratic in $a_\mu$. The
part of the effective action quadratic in the nonet fields 
finally reads
\begin{equation}
{\cal S}=
{1\over 2}\int  \; d^4 x \left(
(\partial_\mu \phi^a-b_\mu^a) {\cal A}_{ab}
(\partial_\mu \phi^b-b_\mu^b)-
\phi^a {\cal B}_{ab} \phi^b+
 \phi^a{\cal C}_{ab}\partial_\mu b_\mu^b \right),
\label{accio}
\end{equation}
where $b_\mu=2 f a_\mu$,  $\phi^a$ ($a=0,1,...,8$) are the fields with
$SU(3)_{R+L}$ quantum numbers and
\begin{equation}
{\cal A}=I+\Delta A\quad,\quad {\cal B}=2 m B (D+\Delta D)
\quad,\quad  {\cal C}=\Delta C \ \ .
\end{equation}
The mass matrix $D$ is non-diagonal already at leading order:
\begin{eqnarray}
D_{11}=D_{22}=D_{33}=1\qquad ,\qquad
D_{44}=D_{55}=D_{66}=D_{77}=1+{x\over 2}\ ,\nonumber\\
D_{88}=1+{2\over 3}x\quad ,\quad
D_{08}=-{{\sqrt 2}\over 3} x\quad ,\quad 
D_{00}=1+{1\over 3} x-{3\over 2}{v_{02}\over mB} \ ,
\end{eqnarray}
with $m=(m_u+m_d)/2$, $x=(m_s-m)/m$.
On the other hand,
the matrix $\Delta A$  brings the next-to-leading order
corrections to the kinetic term,
\begin{eqnarray}
\label{deltaa}
&&\Delta  A_{11}=\Delta A_{22}=\Delta A_{33}=16 {mB\over f^2} L_5
 \ ,\nonumber\\
&&
\Delta A_{44}=\Delta A_{55}=\Delta A_{66}=\Delta A_{77}=
16 {mB\over f^2} L_5\left(1+{x\over 2}\right)  \ , \nonumber \\
&&\Delta  A_{88}=16 {mB\over f^2} L_5\left(1+{2\over 3}x\right)
 \ ,\nonumber \\
&&\Delta A_{08}=-{16 {\sqrt 2}\over 3} {mB\over f^2} L_5 x
 \ ,\nonumber \\
&&\Delta A_{00}=16 {mB\over f^2} L_5
\left(1+{1\over 3} x\right) \  ,
\end{eqnarray}
and $\Delta D$ corrects the mass matrix,
\begin{eqnarray}
&&\Delta D_{11}=\Delta D_{22}=\Delta D_{33}=32 {mB\over f^2} L_8
 \ ,\nonumber\\
&&\Delta D_{44}=\Delta D_{55}=\Delta D_{66}=\Delta D_{77}=
32 {mB\over f^2} L_8\left(1+x+{1\over 4}x^2\right)\nonumber\\
&&\Delta  D_{88}=32 {mB\over f^2} L_8\left(1+{4\over 3}x
+{2\over 3} x^2\right)
 \ ,\nonumber\\
&&\Delta  D_{08}=-{32 {\sqrt 2} \over 3} {mB\over f^2} L_8
x(2+x)+{\sqrt 2} v_{31} x
 \ ,\nonumber\\
&&\Delta  D_{00}=32 {mB\over f^2} L_8 \left(1+{2\over 3} x +{1\over 3}
x^2\right)-
2 v_{31} (3+x) \ .
\end{eqnarray}
$\Delta C$ comes entirely from the $U_A(1)$ anomaly:
\begin{eqnarray}
&&\Delta C_{00}= \frac{3}{2} v_{50}\ .
\end{eqnarray}

From the correlator of two axial currents we can read off the physical masses
and the decay constants, the former from the location of the poles, the latter
from their residues. Following the steps of ref. \cite{GL}, the correlator is
the second derivative of (\ref{accio}) with respect to $a_\mu$ when
evaluated for $a_\mu$ that minimises the effective
action; it boils down to a simultaneous diagonalisation 
of ${\cal A}$ and ${\cal B}$ for the masses,
\begin{equation}
{\cal A}= F^\dagger\ I \ F\nonumber\qquad,\qquad
{\cal B}=F^\dagger \ M^2 \ F\\ \ ,
\end{equation}
whereas the decay constants read 
($P$ is a diagonal index, $\alpha$ is non-diagonal),
\begin{equation}
f_{P\alpha}=f \left((F^\dagger)^{-1} \left( {\cal A}
 + {{\cal C}\over 2 }\right)\right)_{P\alpha}
\end{equation}

Consistently,
we perform a perturbative diagonalization, order by order in the expansion. 
First, we diagonalize to leading order and only $D$ has a non-diagonal part;
it is already
diagonal in $\pi$ and $K$ ( we shall neglect the isospin breaking ),
\begin{eqnarray}
&&2mB=M^2_\pi \nonumber\\
&&2mB\left(1+{x\over 2}\right)=M^2_K \ ,
\label{pik}
\end{eqnarray}
only the 0 and 8 components do mix,
\begin{equation}
2 m B  \  F_0 D F_0^\dagger = M^2\ \ ,
\label{rot}
\end{equation}
with
\begin{equation}
F_0=\pmatrix{\cos \theta_0 &-\sin\theta_0 \cr
\sin\theta_0 &\cos\theta_0\cr}\;.
\label{mrot}
\end{equation}
From (\ref{pik}) we can fix $mB$ and $x$ unambiguously. Upon diagonalization
of (\ref{rot})
we fix $v_{02}$ so as to describe correctly $M^2_{\eta'}$. This implies
\begin{equation}
\label{zerofit}
\theta_0\simeq - 20^o,\qquad \qquad v_{02}\simeq -0.22\  GeV^2\ .
\end{equation}
In this way the prediction for $M^2_\eta\simeq (549\ MeV)^2$ which is only
a 10\% off. This is a known feature of the leading $1/N_c$ 
approximation: it was
proven in ref. \cite{howard} that the ratio $M_\eta^2 / M_{\eta'}^2$ always
comes out too small; the next-to-leading
corrections are thus needed to reconcile the $1/N_c$ expansion with 
the observed masses \cite{santi}.

Let us emphasize that we
had two masses, $M^2_\eta$ and $M^2_{\eta'}$, to fit and just one parameter,
$v_{02}$, to tune. In principle, we had
the choice to fit either one
to its experimental value.
We could  have, instead,  fixed the value of $v_{02}$
in order to describe correctly $M^2_\eta$, obtaining a prediction
for $M^2_{\eta'}$. This choice would give, though, a poor prediction
for $M^2_{\eta'}$ since 
it is the $\eta'$ that gets the bulk of the contribution from 
$v_{02}$ being, therefore, more sensitive to it.

The next-order expressions for the masses 
can also be found in \cite{L}. The $\Delta A$ corrections are absorbed
into a wave-function renormalization and
the $\pi$ and $K$ are still diagonal
\begin{eqnarray}
&& M^2_\pi= 2 m B \left( 1 + 16{m B \over f^2} (2 L_8-L_5)
\right) \ \ ,\nonumber \\
&& M^2_K= 2 m B
\left(1+{x\over 2}\right)\left( 1 + 16{m B \over f^2}
 (2 L_8-L_5)\left(1+{x\over 2}\right)
\right)\ \ ,
\end{eqnarray}
 whereas the 0 and 8 components remain to be diagonalized and
can be written as
\begin{eqnarray}
&& m^2_{88}= {4 M^2_K-M^2_\pi\over 3} +
{4\over 3} (M^2_K-M^2_\pi) \Delta_M\ \ ,
 \nonumber \\
&& m^2_{08}= {-2{\sqrt 2}\over 3}(M^2_K-M^2_\pi)
(1+\Delta_M-\Delta_N)\ \ ,
 \nonumber \\
&& m^2_{00}={1\over 3}(2 M^2_K+M^2_\pi)(1- 2 \Delta_N)
+{2\over 3}(M^2_K-M^2_\pi) \Delta_M - 3 v_{02}\ \ ,
\end{eqnarray}
where
\begin{equation}
\Delta_M={8\over f^2}
(M^2_K-M^2_\pi) (2 L_8- L_5) \ \ , \  \
\Delta_N= 3 v_{31} - {12\over f^2} v_{02} L_5 \ .
\end{equation}
Note the explicit violation of the  Kaplan-Manohar symmetry \cite{KM} 
in $SU(3)_L \otimes SU(3)_R$ -
usually stated as the fact that in the expressions for the masses 
$L_8$ and $L_5$ always come
through the combination $(2 L_8-L_5)$ - 
in $U(3)$ chiral perturbation theory which 
is apparent in $\Delta_N$: this correction brings about a different
dependence, on $L_5$ only, due to the presence of $v_{02}$, i.e., the $U_A(1)$
anomaly.  

Let us now proceed to diagonalise to next-to-leading order.
In order to preserve the form of the above partial result,
we write the transformation matrix as
\begin{equation}
F =
\pmatrix{\cos \theta&-\sin\theta\cr
\sin \theta & \cos\theta\cr}
\ \pmatrix{1+{\Delta A_{88}\over 2}
& {\Delta A_{08}\over 2}\cr
 {\Delta A_{08}\over 2}
&1+{\Delta A_{00}\over 2} } \ \ .
\end{equation}
where we have explicitly separated the rotation
that diagonalizes the kinetic term. Note that
$\theta$ also gets  corrections with respect
to  $\theta_0$. 
For the physical masses we find,
\begin{eqnarray}
M^2_\eta &=& (1-{y\over 3}) M^2_K +{y\over 3} M^2_\pi
- {\sqrt{9+2 y +y^2}\over 3}(M^2_K-M^2_\pi)
+\left( 1 -{9 +y\over 3\sqrt{9+2 y +y^2}}\right)
(M^2_K-M^2_\pi) \Delta_M \nonumber \\
&+&\left(-{1\over 3}(2 M^2_K+M^2_\pi) +
{1\over 3\sqrt{9+2 y +y^2}}(3(2 M^2_K-3 M^2_\pi)
+ y (-2 M^2_K-M^2_\pi))\right)\Delta_N\ \ ,
\end{eqnarray}
where
\begin{equation}
 y\equiv {9\over 2}{v_{02}\over M^2_K-M^2_\pi}\ \ 
\end{equation}
and
\begin{equation}
M^2_\eta+M^2_{\eta'}=
-{2\over 3}(-3+y)M^2_K+{2\over 3}y M^2_\pi+
2 (M^2_K-M^2_\pi) \Delta_M-{2\over 3} (2M^2_K+M^2_\pi)\Delta_N \ ;
\end{equation}
the angle $\theta$ being,
\begin{equation}
\tan 2\theta = {2\sqrt{2}\over 1+y}
\left(1 + {y\over 1+y}\Delta_M-\Delta_N-
{1\over 1+y} {2 M^2_K+M^2_\pi\over M^2_K-M^2_\pi}\Delta_N
\right)\ .
\end{equation}

Let us now turn to the numerical exploitation of these results. The global 
counting of parameters versus data goes as follows: we are left with 
$y$ (which is proportional to $v_{02}$), $\Delta_M$ and $\Delta_N$ to
fix $M^2_\eta$ and $M^2_{\eta'}$. Thus, no prediction can be
made. Yet, if we fix a rotation angle, then there is a fit
for all three parameters. 
The results for angles which are close to the zeroth order 
$\theta_0$ are displayed in the following table:

\begin{table}[h]
\begin{center}
\begin{tabular}{|c|c|c|c|}
\hline
$\theta$&$\Delta_M$&$\Delta_N$&$-{v_{02}\over f^2}$\\
\hline
$-18^o$&$0.113$&$0.270$&$30.8$\\
\hline
$-20^o$&$0.156$&$0.220$&$29.3$\\
\hline
$-22^o$&$0.203$&$0.178$&$27.9$\\
\hline
$-24^o$&$0.254$&$0.143$&$26.5$\\
\hline
\end{tabular}
\end{center}
\label{t1}
\end{table}

The best fit, if $U(3)$ chiral perturbation theory 
is to make sense, is expected to be near the zeroth order result.
One could  play with different convergence criteria to
argue in favor of a given angle. A nice feature of our table is that
$\Delta_M$ and $\Delta_N$ move in different directions as we
change the mixing angle. This leads to a minimization of corrections
around $\theta\sim -20^o$. Moreover, one may select
the optimal $\theta$ where the masses of $\eta$ and $\eta'$
receive the smallest correction
in the sense that corrections proportional to $\Delta_M$ and $\Delta_N$ 
tend to cancel.
Interestingly enough, we
find it to be very close to $-20^o$ as well.

Of course none of these arguments is a substitute for a global fit to data.
However, our results agree in hinting 
at a mixing angle close to $-20^o \geq \theta \geq -21^o$, 
which correspond to values of $\Delta_M$ and $\Delta_N$ compatible with
those given in \cite{L} ($\Delta_M\simeq 0.18$, $\Delta_N\simeq 0.24$).
We stress that since from this point of view the angle $\theta$ 
is the only variable, the values of $\Delta_M$ and $\Delta_N$ are correlated.

At this point let us comment on the Gell-Mann-Okubo formula, with corrections
from chiral perturbation theory. The violation of the original
formula is related to $\Delta_M$,
\begin{equation}
4 M^2_K- M^2_\pi-3 M^2_\eta-3 \sin^2\theta \left(
M^2_{\eta'}-M^2_\eta\right) = -4 (M^2_K- M^2_\pi) \Delta_M \ .
\label{GO}
\end{equation}
The leading order equation, thus, corresponds to setting the 
r.h.s. to zero. At this order, $\theta\sim-10^0$ in order to fulfil the
equation. This is not the procedure we chose
in Eq.\ref{zerofit}, as we took $M_{\eta'}^2$
to fix  $\theta\sim -20^0$ at leading order. Eq. (\ref{GO})
thus provides a different way of 
fitting the mixing, which is strongly corrected by $\Delta_M$.
Of course, at next-order, the best fit at e.g. $\Delta_M\sim 0.156$
is $\theta\sim -20^0$ as well.

\bigskip

We may also use our analytical results to get the form of
the decay constants for the whole nonet. The diagonal elements
of relevance are given by the combination
\begin{equation}
f_{physical}= f
\pmatrix{\cos \theta&-\sin\theta\cr
\sin \theta & \cos\theta\cr}
\pmatrix{1+{\Delta A_{88}\over 2}
& {\Delta A_{08}\over 2}\cr
 {\Delta A_{08}\over 2}
&1+{\Delta A_{00}\over 2}+{\Delta E_{00}\over 2}\cr}
\pmatrix{\cos \theta&\sin\theta\cr
 -\sin \theta & \cos\theta\cr} \ .
\end{equation}
This expression yields
\begin{eqnarray}
&&f_\pi= f \left(1 + 4 {L_5 \over f^2}M^2_\pi\right) \ ,
\nonumber \\
&&f_K= f \left(1 + 4 {L_5\over f^2} M^2_K\right) \ ,
\nonumber \\
&&f_\eta= f \left(1 + 4 {L_5\over f^2}\left(M^2_K-
{M^2_K-M^2_\pi\over 3} {9+y\over \sqrt{9+2 y +y^2}}\right)
+ \frac{3}{8}v_{50} \left( 1+ \frac{1+y}{\sqrt{9+2 y +y^2}}\right)\right)
\ ,\nonumber \\
&&f_{\eta'}= f \left(1 + 4 {L_5\over f^2}\left(M^2_K+
{M^2_K-M^2_\pi\over 3} {9+y\over \sqrt{9+2 y +y^2}}\right)
+ \frac{3}{8}v_{50} \left( 1- \frac{1+y}{\sqrt{9+2 y +y^2}}\right)\right)
\ .\label{f}
\end{eqnarray}
Note that in the exact $SU(3)_{R+L}$ limit of quarks degenerate in mass
there is no mixing and one finds
\begin{eqnarray}
&&f_{\eta}\rightarrow f\left(1+4{L_5\over f^2}
{4 M^2_K-M^2_\pi\over 3}\right)\ ,\nonumber \\
&&f_{\eta'}\rightarrow f\left(1+4{L_5\over f^2}
{2 M^2_K+ M^2_\pi\over 3} + \frac{3}{4}v_{50}\right)\ ,
\end{eqnarray}
which is the $y \rightarrow -\infty$ limit of (\ref{f}), as expected.
\bigskip

Let us compare these results with data.
We can take $f_\pi$ and
$f_K$  to fix $f$ and $L_5$, ($f_\pi=92.4$  MeV,
$f_K=1.223 f_\pi$)
\begin{equation}
f=90.8 {\rm MeV}\qquad , \qquad L_5=2.0 \ 10^{-3}\ .
\end{equation}
This is the same value  as obtained 
for $L_5^{[SU]\, r}$
in \cite{GL} for it is extracted from the same source.
 In our approximation 
the issue of the running of the $L_5$, $L_8$ cannot 
be addressed because the 
loop corrections are dropped altogether. 
As a matter of principle one expects
$L_5 \sim L_5^{[SU]}$ because integrating out the $\eta'$
does not give any contribution to 
$L_5^{[SU]}$, $L_8^{[SU]}$ at tree level, except for $L_7^{[SU]}$. 

We are left with  $f_\eta$ and $f_{\eta'}$
to be described by $y$ (again, proportional to $v_{02}$) and
 $v_{50}$. Assuming that convergence criteria fix
a rotation angle around $\theta\sim -20^o$, this leaves
room for a  fit of $v_{50}$ and a prediction. The values of 
$\eta$ and $\eta'$ decay constants
are, though, poorly determined. Indeed,
the experimental values of  $f_\eta$ and $f_{\eta '}$ 
 \cite{pdb96} have large errors:

\begin{eqnarray}
0.943 \leq& \frac{f_\eta}{f_\pi} &\leq 1.091 \ ,\\  \nonumber
0.912 \leq& \frac{f_{\eta'}}{f_\pi} &\leq 1.015 \ .\nonumber
\end{eqnarray}

Let us present the
output of our formulae as a function of the mixing angle $\theta$.
At each given $\theta$, the experimental values of $\frac{f_\eta}{f_\pi}$ 
allow for a range of values 
of the parameter $v_{50}$ and of $\frac{f_{\eta'}}{f_\pi}$. 
We again display the results for angles which are
close to the zeroth order $\theta_0$.

\begin{table}[h]
\begin{center}
\begin{tabular}{|c|c|c|}
\hline
$\theta$&
$- v_{50} \; min/max$&
${f_{\eta'}\over f_\pi} \; min/max $\\
\hline
$-20^o$&$2.293/0.601$&$-0.218/0.902$\\
\hline
$-22^o$&$1.775/0.365$&$0.171/1.080$\\
\hline
$-24^o$&$1.394/0.197$&$0.457/1.205$\\
\hline
\end{tabular}
\end{center}
\end{table}

A word of caution should be added as regards the use of the $f_{\eta'}$ value,
as made in this paper. As pointed out by Shore and Veneziano
\cite{ShV}, from the decay rate of $\eta' \to \gamma \gamma$ one cannot
obtain the value of $f_{\eta'}$ because the singlet axial current $A^{(5)}_\mu$
is not conserved in the chiral limit, due to the $U_A(1)$ anomaly, which makes
$f_{\eta'}$ from 
$\langle 0 |A^{(5)}_\mu |\eta'(k) \rangle = i k_\mu f_{\eta'}$ 
depend on a subtraction point and thus not be observable. Of
course this dependence is $1/N_c$ suppressed because in the large-$N_c$ limit
the anomaly is a subleading effect. The relation between our 
$f_{\eta'}$ and what is measured in $\eta' \to \gamma \gamma$ and given in 
\cite{pdb96}, \cite{Bij} can be worked out and ammounts
to include one further term in the effective action
coupled to an external electromagnetic source \cite{Nos}.
This would still bring in a new unkown constant and we would face the
problem of having one more constant to fit than measured constants available.
Nevertheless, we overcome this shortcoming by
using the criterium of {\it minimum sensitivity} that minimizes
the size of the corrections to a given order and conclude that the combined 
expansion (\ref{E2}), within the framework of $U_L(3) \otimes U_R (3)$
chiral perturbation theory, is able to accommodate very naturally the observed
values of masses and decay constants $f_\pi$, $f_K$. The big uncertainty
in $f_\eta$ makes the method inconclusive for $f_{\eta '}$ (which, strictly
speaking, is not experimentally known), as can be seen from the last table.
If data from $\eta' \to \gamma \gamma$ were more precise, we could with our
method pursue to determine the new contant that is involved in the process,
and how well it would accommodate in the framework of the combined expansion.

We finish by quoting the values of all the parameters in one batch, as
functions of the mixing angle in the range $20^o < \theta < 24^o$: 
\begin{eqnarray}
0.15  \leq& {\Delta}_M & \leq 0.26  ,\\ \nonumber
0.980  \leq & {2 m B\over M^2_\pi} & \leq  0.988 , \\ \nonumber
18.3 \leq & x & \leq  20.9  ,\\ \nonumber
-4.7 \leq & y & \leq -4.2  ,\\ \nonumber
26 \leq & -v_{02}/f^2 & \leq 29   ,\\ \nonumber
0.14 \leq & {\Delta}_N & \leq 0.22  ,\\ \nonumber
1.35 \ 10^{-3} \leq& L_8 &\leq  1.57 \ 10^{-3}  ,\\ \nonumber
-0.164 \leq& v_{31}& \leq  -0.161  .\nonumber
\end{eqnarray}
\vspace*{1cm}

We should like to add that after this work was finished there appeared the
document of ref. \cite{HL} in which the author reports
some results, as yet unpublished, on the same issue that concerns our study.

\vspace*{1cm}
\section{Acknowledgments}

Financial support from CICYT, contract AEN95-0590,
and from CIRIT, contract GRQ93-1047 are
acknowledged. J.I. L. acknowledges the Benasque Center for Physics
where part of this work was completed.
P.H.-S. acknowledges a Grant from the {\it Generalitat de Catalunya}.
J.T. acknowledges the Theory Group at CERN for the hospitality extended
to him.

\end{document}